\documentclass[twoside]{article}
\usepackage{fleqn,espcrc2}

\usepackage{epsfig}

\usepackage[figuresright]{rotating}

\newcommand{\AmS}{{\protect\the\textfont2
  A\kern-.1667em\lower.5ex\hbox{M}\kern-.125emS}}

\newcommand{\Sign}{\mathrm{Sign}}

\title{Fermion Cluster Algorithms }
\author{Shailesh Chandrasekharan
	\address{Department of Physics, Duke University,
        Durham, NC 27708}%
        \thanks{This work was supported in part by US department of
		energy grant DE-FG02-96ER40945.}
     }  
\begin{document}

\begin{abstract}
Cluster algorithms have been recently used to eliminate sign problems 
that plague Monte-Carlo methods in a variety of systems. In particular 
such algorithms can also be used to solve sign problems associated with 
the permutation of fermion world lines. This solution leads to the 
possibility of designing fermion cluster algorithms in certain cases. 
Using the example of free non-relativistic fermions we discuss the 
ideas underlying the algorithm. 
\end{abstract}
\maketitle

\section{INTRODUCTION}

Fermions are quite difficult to deal with in Monte-Carlo methods. 
The main problem is the Pauli principle which introduces negative 
Boltzmann weights when fermions are treated physically as particles 
traveling in time. Until a few months ago the only known approach 
was to integrate them out and hope that the remaining problem is described
by a positive Boltzmann weight. This approach has been successful in a few 
cases of physical interest like lattice QCD at zero chemical potential
and zero vacuum angle. Unfortunately, even in these cases the 
Hybrid-Monte-Carlo methods slow down dramatically due to critical 
slowing down in the chiral limit \cite{Sha98}.

Recently a new class of fermion algorithms have been discovered
where the fermions are treated as physical particles traveling in time
\cite{Cha99.1}. Such world line algorithms had been suggested in the 
past, however no solution to the fermion sign problem was 
found\cite{Uwe93,Tro97}. We have been able to solve this problem using 
cluster algorithms in a limited class of models. The progress is due to 
a better understanding of the relation between the topology of clusters and
the effect of their flip on the fermion permutation sign. In certain cases, 
this knowledge can be used to completely eliminate the fermion sign problem 
and at the same time gain from the ability of cluster algorithms to beat 
critical slowing down. Thus, when successful these algorithms lead to the 
most ideal fermion algorithms discovered so far.

\section{SIGN PROBLEM AND SOLUTION}

The fermionic cluster algorithms that we have discovered are based on the 
cluster algorithm for a bosonic quantum spin-1/2 model. It is well known
that a site with a fermion can be identified with a spin ``up'' state and
an empty site with a spin ``down'' state up to sign factors that arise due 
to the Pauli principle. For every spin configuration the spin ``up'' states
can be used to track fermion world lines which are closed in Euclidean time 
and describe a permutation of fermions. The sign of this permutation is 
exactly the same as the product of sign factors that arise from the Pauli 
principle. Hence it is always possible to absorb the physics of the Pauli 
principle into the sign of the Boltzmann weight of the spin configuration. 
Mathematically, the fermionic partition $Z_f$ can be written as
\begin{equation}
Z_f =  \sum_{[{\cal C}]}\; \Sign[{\cal C}]\;W_b[{\cal C}],
\;\;Z_b = \sum_{[{\cal C}]}\;  W_b[{\cal C}],
\end{equation}
where $Z_b$ is the partition function of the quantum spin model written as 
a sum over configurations ${\cal C}$, defined by a set 
${\cal C}_i,\;i=1,2,...N_C$ of connected spin clusters, with a positive 
Boltzmann weight $W_b[{\cal C}]$. In general
the sign factor in $Z_f$ is a product of the fermion permutation sign 
$\Sign_f[{\cal C}]$, discussed above and other local sign factors
$\Sign_b[{\cal C}]$, that may be necessary to relate the fermionic and the 
bosonic models. Further, the existence of a cluster algorithm implies 
that the weight $W_b[{\cal C}]$ remains the same if all the spins of any 
connected spin-cluster are flipped. On the other hand the $2^{N_C}$ degenerate 
configurations, obtained by flipping the clusters, can have different sign 
factors $\Sign[{\cal C}]$. 

Although the above method of writing the fermionic model in terms of 
clusters of a bosonic model is well known, the freedom in choosing the 
bosonic weights $W_b[{\cal C}]$ and local bosonic sign factors 
$\Sign_b[{\cal C}]$ had not been exploited until now. We have discovered that 
it is always possible to be clever in using this freedom so that the 
connected spin-clusters contribute independently to the sign of the 
configuration, i.e., the overall sign can always be written as 
$\Sign[{\cal C}] = \prod_{i=1}^{N_C}\;\; \Sign({\cal C}_i)$,
where $\Sign({\cal C}_i)$ is the sign associated with a connected cluster 
of spins. If $\Sign[{\cal C}_i]$ changes when the spins are flipped
the cluster is called a {\sl meron}\footnote{This word was originally
used in \cite{Bie95} to describe clusters with the same property in a
classical $O(3)$ model.}. Thus, meron clusters identify two spin 
configurations with the
same weight but opposite signs and hence only non-meron clusters contribute
to the partition function. It is always possible to include a Metropolis
decision during the cluster formation process to suppress meron clusters in
a controlled way. In certain models the spins within any connected
cluster can always be flipped to a reference pattern 
${\cal C}_i^{\mathrm{ref}}$ such that $\Sign[{\cal C}_i^{\mathrm{ref}}]=1$. 
In such cases the average of $\Sign[{\cal C}]$ under all the $2^{N_C}$ 
flips of connected spin-clusters is $1$ in the zero meron sector and this
solves the sign problem completely.

\section{MODELS AND ALGORITHMS}

There are a variety of models that can be solved using the above 
ideas. Here we discuss how these ideas can be applied to solve
free non-relativistic fermions on a d-dimensional hyper-cubic lattice
described by the Hamiltonian,
\begin{equation}
\label{nrferm}
H \!=\! \sum_{x,i} \left( 
n_x + n_{x+\hat{i}} - [c_x^+ \;c_{x+\hat{i}} + c_{x+\hat{i}}^+ \;c_x]\right),
\end{equation}
where $n_x = c_x^+ c_x$ is the fermionic occupation number and 
$c^\dagger_x$ and $c_x$ are creation and annihilation operators.
This model was originally considered in \cite{Uwe93}. However, due
to a wrong choice of $W_b[{\cal C}]$ and $\Sign_b[{\cal C}]$,
even this simple model appeared intractable numerically. Here we
show how a different choice of these factors solves the problem 
completely. Following \cite{Uwe93} we construct the partition function 
$Z_f = \mathrm{Tr}[\exp(-\beta H)]$ by discretizing the Euclidean time 
axis into $2d\times M$ steps such that at a given time slice each spin 
interacts with only one neighboring spin. Thus, the Boltzmann weight of any 
configuration of fermion occupation numbers is a product of two-spin 
transfer matrix elements up to the global fermion permutation sign. 
Figure 1 illustrates a typical configuration in the path integral with 
the shaded regions representing the two spin interactions. 
\begin{figure}[ht]
\vspace{-4cm}
\begin{center}
\hbox{
\hspace{2cm}
\epsfig{file=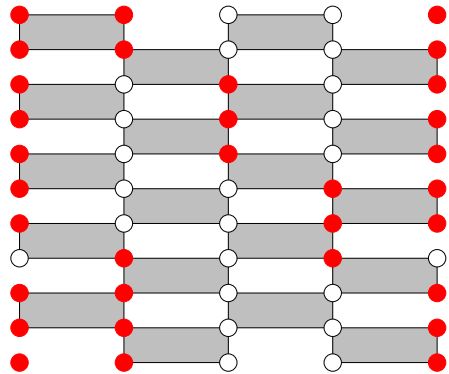,
width=4.5cm,angle=0,
bbllx=50,bblly=50,bburx=225,bbury=337}
}
\hspace{0.5cm}
\caption{ A typical spin configuration whose Boltzmann weight
(up to an overall constant) is 
$\mathrm{e}^{-3\epsilon}\mathrm{e}^{3\epsilon}[\cosh(\epsilon)]^{10}
[\sinh(\epsilon)]^4$
and the fermion permutation sign is $-1$. Here $\epsilon = \beta/M$.}
\end{center}
\end{figure}
\vspace{-1cm}

In order to construct the cluster algorithm we next introduce bond 
variables in addition to spin variables to describe connected 
spin-clusters and find transfer matrix elements for these new types 
of connected spin configurations such that the partition function remains
the same. A given spin configuration can represent many spin-cluster 
configurations all of which have the same global fermion permutation sign. 
If we allow the transfer matrix elements of these new configurations to 
be negative there is a lot of freedom in choosing the weights and signs. 
Figure 2., illustrates a particular choice such that summing over the 
bond variables reproduces the weights of spin configurations. An 
interesting feature of the spin connection rules of figure 2 is that 
all spins in a connected cluster are of the same type. Further, the 
negative sign associated with the cross bond configuration with all 
``up'' spins is an extra local negative sign that can be absorbed into 
$\Sign_b[{\cal C}]$, where as the global fermion permutation sign
is absorbed into the factor $\Sign_f[{\cal C}]$.
\vspace{-5.5cm}
\begin{figure}[ht]
\begin{center}
\hbox{
\hspace{1cm}
\epsfig{file=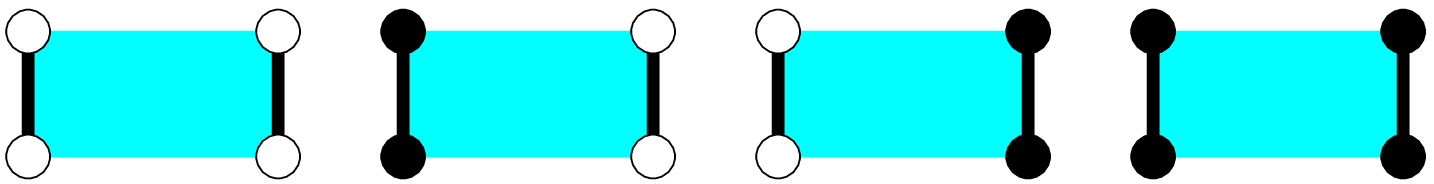,
width=3.0cm,angle=0,
bbllx=50,bblly=50,bburx=225,bbury=337}
}
\vspace{1cm}
\hbox{
\hspace{0.7cm}
(a) \hspace{1.2cm} (b) \hspace{1.2cm} (c) \hspace{1.2cm} (d)
}
\vspace{-4cm}
\hbox{
\hspace{1cm}
\epsfig{file=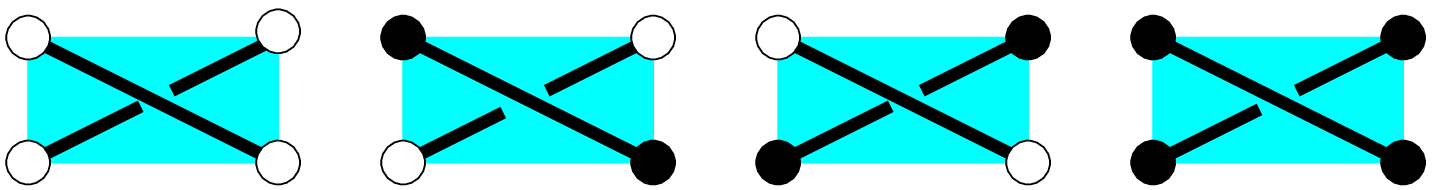,
width=3.0cm,angle=0,
bbllx=50,bblly=50,bburx=225,bbury=337}
}
\vspace{1cm}
\hbox{
\hspace{0.7cm}
(e) \hspace{1.2cm} (f) \hspace{1.2cm} (g) \hspace{1.2cm} (h)
}
\vspace{-0.5cm}
\caption{ Transfer matrix elements of spin-cluster configurations.
The weights of (a), (b), (c) and (d) are $\cosh(\epsilon)$, 
(e), (f) and (g) are $\sinh(\epsilon)$ and (h) is $-\sinh(\epsilon)$. }
\end{center}
\end{figure}
\vspace{-1.3cm}
The model described by the magnitude of the weights of figure 2, is the
spin-1/2 ferromagnetic Heisenberg model and can be updated using a cluster 
algorithm. Remarkably, $\Sign[{\cal C}]=\Sign_f[{\cal C}]\;\Sign_b[{\cal C}]$ 
has all the desirable properties described in the previous section. The 
sign of a connected spin cluster is set to $1$ if it is a cluster of 
``down'' spins or if the temporal winding of the cluster is odd. Otherwise 
the sign of that cluster is $-1$. Thus, clusters with an even temporal 
winding are merons.  

In summary we have shown that clusters generated with the cluster algorithm 
of the ferromagnetic quantum spin-1/2 Heisenberg model, can also describe 
free non-relativistic fermions whose Hamiltonian is given in 
eq.(\ref{nrferm}), if we throw away clusters with an even temporal winding. 
In order to demonstrate the correctness of this observation we have 
calculated the two point fermion Greens function defined as
\begin{equation}
G(x,y;t) = \frac{1}{Z_f} \mathrm{Tr}\left( \mathrm{e}^{[-(\beta-t)H]}\; 
c_x \mathrm{e}^{[-tH]}\; c_y^+\right),
\end{equation}
on a $4\times 4\times 4$ lattice at $\beta = 1$ and $M=16$. Figure 3 shows 
this function in momentum space for $\vec{p}=(0,0,0)$ and $\vec{p}=(\pi,0,0)$. 
Evidently, the exact zero mode of $H$ at zero momentum does not lead to 
any complications, unlike conjugate gradient methods.

It is possible to extend the model to include short range repulsive 
interactions. In addition, the above ideas are also applicable in a
variety of models with a rich phase structure. One such model has a 
finite temperature chiral phase transition and was studied extensively in 
\cite{Cha99.2}. The cluster algorithm of the anti-ferromagnetic Heisenberg 
model plays an important role there. Remarkably one can work directly in the 
chiral limit and no critical slowing down is observed. 
\begin{figure}[ht]
\vspace{1cm}
\begin{center}
\hbox{
\hspace{0.2cm}
\epsfig{file=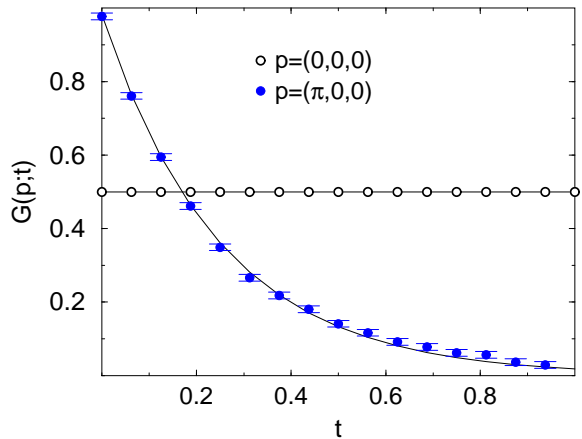,
height=4.3cm,angle=0,
bbllx=50,bblly=50,bburx=225,bbury=337}
}
\vspace{-1cm}
\hspace{0.5cm}
\caption{ Two point correlation function in momentum space. The solid
lines represent exact results.}
\end{center}
\end{figure}
\vspace{-1cm}
All this shows that fermion cluster algorithms can provide a very
elegant method to solve fermionic field theories numerically.

I wish to thank Uwe Wiese for his collaboration and many discussions at 
various stages of this work.


\begin{thebibliography}{9}

\bibitem{Sha98} S. Sharpe, hep-lat/9811006.

\bibitem{Cha99.1} S. Chandrasekharan and U.-J. Wiese, cond-mat/9902128.

\bibitem{Uwe93} U.-J. Wiese, Phys. Lett. B311 (1993) 235.

\bibitem{Tro97} B. Ammon et. al., Phys. Rev. B58, 4304.

\bibitem{Bie95} W. Bietenholz et. al., Phys. Rev. Lett. 75 (1995) 4524.

\bibitem{Cha99.2} S. Chandrasekharan et. al., hep-lat/9806232
		
\end{thebibliography}
\end{document}